\documentclass[prl,twocolumn,superscriptaddress]{revtex4}
\usepackage{times}
\usepackage{graphicx}
\usepackage{latexsym,amsmath,amssymb,bm,euscript}
\usepackage{color}

\def\up{\uparrow}
\def\dn{\downarrow}

\def\ET{{$\kappa$-(ET)$_2$Cu$_2$(CN)$_3$}}
\def\dmit{{EtMe$_3$Sb[Pd(dmit)$_2$]$_2$}}

\def\hx{{\hat{x}}}
\def\hy{{\hat{y}}}
\def\hxy{{\hx + \hy}}
\begin{document}

\title{Spin Bose-Metal and Valence Bond Solid phases in a spin-1/2 model with ring exchanges on a four-leg triangular ladder}

\author{Matthew S. Block}
\affiliation{Department of Physics, University of California, Santa Barbara, California 93106, USA}

\author{D. N. Sheng}
\affiliation{Department of Physics and Astronomy, California State University, Northridge, California 91330, USA}

\author{Olexei I. Motrunich}
\affiliation{Department of Physics, California Institute of Technology, Pasadena, California 91125, USA}

\author{Matthew P. A. Fisher}
\affiliation{Department of Physics, University of California, Santa Barbara, California 93106, USA}
\affiliation{Department of Physics, California Institute of Technology, Pasadena, California 91125, USA}

\date{\today}

\begin{abstract}
We study a spin-1/2 system with Heisenberg plus ring exchanges on a four-leg triangular ladder using the density matrix renormalization group and Gutzwiller variational wave functions.  Near an isotropic lattice regime, for moderate to large ring exchanges we find a spin Bose-metal phase with a spinon Fermi sea consisting of three partially filled bands.  Going away from the triangular towards the square lattice regime, we find a staggered dimer phase with dimers in the transverse direction, while for small ring exchanges the system is in a featureless rung phase.  We also discuss parent states and a possible phase diagram in two dimensions.
\end{abstract}

\maketitle

%%%%%%%%%%%%%%%%%%%%%%%%%%%%%%%%%%%%%%%%%%%%%%%%%%%%%%%%%%%%%%%%%%%%%

In a wide class of crystalline organic Mott insulators it is possible to tune from the strongly correlated insulating state into a metallic state.  At ambient pressure such ``weak Mott insulators" are perched in close proximity to the metal-insulator transition.  The residual electronic spin degrees of freedom constitute a novel quantum system and can exhibit a myriad of behaviors such as antiferromagnetic (AF) ordering or a valence bond solid (VBS).  Particularly exciting is the possibility that the significant charge fluctuations in a weak Mott insulator frustrate the magnetic or other ordering tendencies, resulting in a quantum spin liquid.  This appears to be realized in two organic materials \cite{Shimizu03, Kurosaki, SYamashita, MYamashita, Itou08, MYamashita2010, Itou2010} \ET\ and \dmit, both quasi-two-dimensional (2D) and consisting of stacked triangular lattices.  Thermodynamic, transport, and spectroscopic experiments point towards the presence of many gapless excitations in the spin-liquid phase of these materials.

\begin{figure}[t]
\centerline{\includegraphics[width=2.75in]{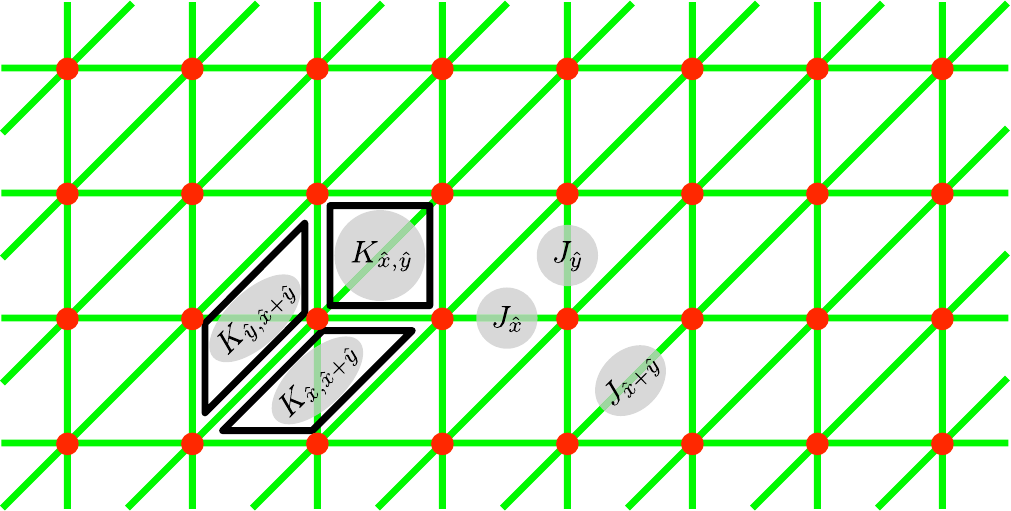}}
\caption{
(color online).  Picture of the Heisenberg plus ring Hamiltonian on the four-leg ladder showing different two-spin and four-spin couplings.  The isotropic model is defined by $J_\hx = J_\hy = J_\hxy = J$, $K_{\hx,\hy} = K_{\hx,\hxy} = K_{\hy,\hxy} = K$.  We also study a broader phase diagram interpolating between the triangular and square limits by decreasing $J_\hxy$ [with appropriate scaling of the ring couplings $K_{\hx,\hxy} = K_{\hy,\hxy} = \left(J_\hxy/J_\hx\right) K_{\hx,\hy}$].  The ladder has periodic boundary conditions in both directions.
}
\label{fig:model}
\end{figure}

The triangular lattice Hubbard model \cite{McKenzie, Powell2010, Shimizu03, Tocchio2010} is commonly used to describe these materials.  At half filling the Mott metal-insulator transition can be tuned by varying the single dimensionless parameter, the ratio of the on-site Hubbard $U$ to the hopping strength $t$.  On the insulator side at intermediate $U/t$, the Heisenberg spin model should be augmented by multispin interactions \cite{McDonald, ringxch, SSLee, Senthil_Mott, Grover2010, Yang2010}, such as four-site ring exchanges (see Fig.~\ref{fig:model}), which mimic the virtual charge fluctuations.  Accessing a putative gapless spin liquid in 2D in such models poses a theoretical challenge.

Slave particle approaches provide one construction of gapless spin liquids and predict spin correlations that decay as power laws in space, oscillating at particular wave vectors.  In the so-called ``algebraic spin liquids" \cite{WenPSG, Rantner02, Hermele_U1, LeeNagaosaWen} these wave vectors are limited to a finite discrete set, often at high symmetry points in the Brillouin zone.  However, the singularities can also occur along surfaces in momentum space, as they do in a ``spinon Fermi sea'' spin liquid speculated for the organic materials \cite{ringxch, SSLee, Senthil_Mott}.  We will call such a phase a ``spin Bose-metal'' (SBM) state \cite{SBM_Solvay, 2legSBM} to emphasize that it has metal-like properties for spin and energy transport while the spin model is bosonic in character.

It should be possible to access an SBM phase by systematically approaching 2D from a sequence of quasi-1D ladder models \cite{2legDBL, SBM_Solvay, 2legSBM}.  On a ladder the quantized transverse momenta cut through the 2D surface, leading to a quasi-1D descendant state with a set of low-energy modes whose number grows with the number of legs. These quasi-1D descendant states can be analyzed in a controlled fashion using numerical and analytical approaches.

{\it Heisenberg plus ring on a four-leg triangular ladder.}---Pursuing this idea, we consider a spin-1/2 system with Heisenberg and four-site ring exchanges,
\begin{eqnarray}
\hat{H}=\sum_{\langle ij\rangle}2J_{ij}\vec{S}_{i}\cdot\vec{S}_{j}+\sum_{\mathrm{rhombi}}K_{\mathrm{P}}\left({\cal P}_{1234}+\mathrm{H.c.}\right).
\end{eqnarray}
An earlier exact diagonalization (ED) work \cite{LiMing} on the isotropic 2D triangular lattice found that $K > 0.1J$ destroys the 120$^\circ$ AF order.  A subsequent variational study \cite{ringxch} suggested the spin Bose-metal phase for moderate to large $K$.  A recent work pursued this model on a two-leg zigzag ladder \cite{Klironomos, 2legSBM} combining density matrix renormalization group (DMRG), variational Monte Carlo (VMC), and Bosonization approaches, and argued that it realizes a quasi-1D descendant of the SBM phase: a remarkable 1D quantum phase with three gapless modes and power law spin correlations at incommensurate wave vectors that are the fingerprints of the parent 2D phase.

The two-leg ladder is still far from 2D.  We take a significant step and study the model on a four-leg ladder.  We first consider the case where all nearest neighbor bonds have the same coupling $J$ and all rhombi have the same coupling $K$; thus there is a single parameter $K/J$.

We study the model numerically using DMRG/ED combined with VMC calculations.  All calculations use periodic boundary conditions.  The DMRG calculations keep $m=3600$-$5000$ states per block \cite{White_dmrg1, White_dmrg2, Schollwock} to ensure accurate results, and the density matrix truncation error for our systems is of the order of $10^{-5}$ (typical relative error for the ground-state energy is $10^{-3}$ or smaller).  Information about the state is obtained by measuring spin, dimer, and (scalar) chirality structure factors.

\begin{figure}[t]
\centerline{\includegraphics[width=\columnwidth]{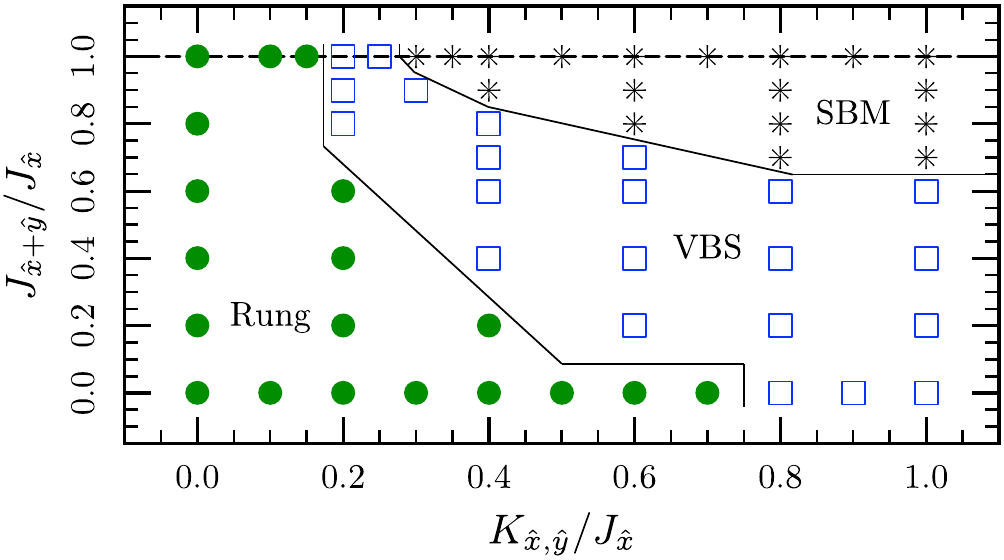}}
\caption{
(color online). Phase diagram of the Heisenberg plus ring model on the four-leg ladder interpolating between the triangular $J_\hx = J_\hy = J_\hxy = 1$ and square $J_\hx = J_\hy = 1, J_\hxy = 0$ limits. 
The horizontal axis is the ring coupling $K_{\hx,\hy}$ while the vertical axis is the diagonal coupling $J_\hxy$, cf.~Fig.~\ref{fig:model}; the other ring couplings are obtained according to Eq.~(\ref{scaledKs}).
}
\label{fig:phased}
\end{figure}

The phase diagram from such a study using $12 \times 4$ and $18 \times 4$ ladders can be seen in Fig.~\ref{fig:phased}; the isotropic case is the horizontal cut at $J_\hxy/J_\hx = 1$.  For small $K/J \leq 0.15$ the system is in a rung phase, whose caricature can be obtained by allowing $J_\hy \gg J_\hx, J_\hxy$ where the rungs effectively decouple.  This phase is gapped and has only short-range correlations.  In the model with isotropic couplings the rungs have rather strong connections: we find that the $\hx$ and $\hxy$ bonds have more negative Heisenberg energies than the $\hy$ bonds.  Nevertheless, the data suggest that the system is in a featureless gapped phase.  A further test is provided by increasing $J_\hy$ from the isotropic case, and we indeed observe a smooth evolution in all measurements towards the strong rung phase.

\begin{figure}[t]
\centerline{\includegraphics[width=\columnwidth]{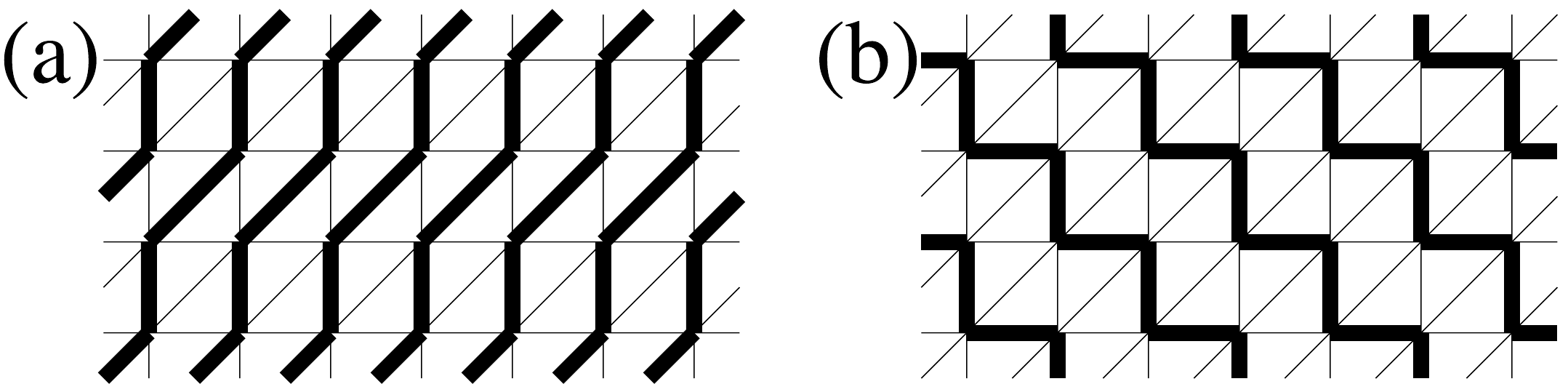}}
\caption{
(a) Symmetry breaking pattern found in DMRG on the isotropic system at $K/J = 0.2$-$0.25$.  (b) State degenerate with (a) in the presence of $\hx \leftrightarrow \hxy$ symmetry.  Both (a) and (b) can be viewed as triangular VBS states with dimers on the $\hy$ bonds but different column orientations.  The staggered patterns on the $\hxy$ and $\hx$ bonds correspondingly are expected on the triangular lattice and follow a rule that each triangle contains only one strong bond.  Upon going to the square limit by decreasing $J_\hxy$, we find state (b), which connects to a staggered $\hy$-dimer state.
}
\label{fig:VBS}
\end{figure}

Near $K/J = 0.2$-$0.25$, the DMRG ground state breaks translational symmetry.  The pattern obtained on both the $12 \times 4$ and $18 \times 4$ systems is illustrated in Fig.~\ref{fig:VBS}(a).  This state has strong $\hy$ bonds forming columns along the ladder direction and strong $\hxy$ bonds in the connecting arrangement.  Note that we also expect a degenerate state depicted in Fig.~\ref{fig:VBS}(b), since the $\hx$ and $\hxy$ directions are equivalent on the isotropic ladder.  The states shown in Fig.~\ref{fig:VBS} are a subset of possible VBS states on the isotropic 2D triangular lattice, and the selection must be due to the finite transverse size.  The selection of (a) in the DMRG must be due to symmetry breaking terms that exist in the way it is building up the multileg system.  Such terms are tiny and translationally invariant ground states are obtained for all other phases without intrinsic degeneracy (we also verified that the DMRG obtained identical results to the ED for $8 \times 4$ systems).

{\it SBM phase.}---For $K/J \geq 0.3$, we do not find any pattern of bond ordering in real space and no indication of Bragg peaks in the dimer or chiral structure factors.  The correlation functions are also markedly different from the rung phase at small $K$.  The $12 \times 4$ and $18 \times 4$ systems remain in essentially the same state for a range of control parameters $0.3 \leq K/J \leq 1$.  Thus, a putative spin-liquid phase is established based on finite-size analysis of the DMRG results.  Spin and dimer correlations are rather extended in real space and show complex oscillations.  The momentum space structure factors allow a more organized view and show many features that, remarkably, can be manifested by simple variational wave functions for the SBM phase.

\begin{figure}[t]
\centerline{\includegraphics[width=\columnwidth]{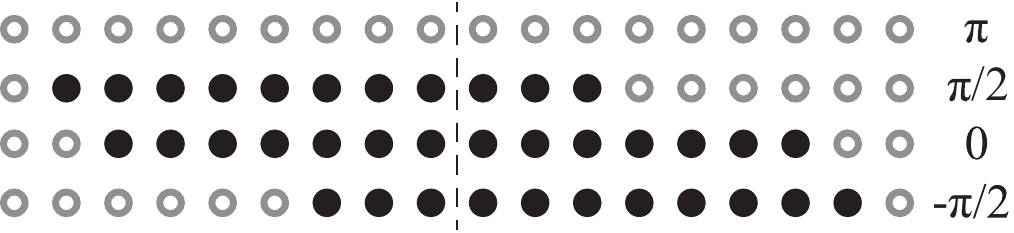}}
\caption{
``Spinon Fermi sea'' of the optimal VMC state for $K/J \geq 0.3$ on the $18 \times 4$ ladder.  Circles denote single-particle orbitals in ${\bm k}$ space;  $\up$ and $\dn$ spinons occupy orbitals shown with filled circles.  There are three partially filled bands and we refer to the resulting state as SBM-3. The dotted line indicates $k_x = 0$, where we are implicitly using antiperiodic boundary conditions in the horizontal direction.  This choice, as well as the location of $k_x=0$, is actually arbitrary as the Gutzwiller-projected wave function is invariant under a global shift of the Fermi sea. 
}
\label{fig:VMCstate}
\end{figure}

To this end, we perform a VMC study using spin-singlet trial wave functions that can be viewed as Gutzwiller projections of spinon hopping mean field states.  More systematically, we vary directly the shape of the ``spinon Fermi sea'' in the momentum ${\bm k}$ space.  There are four transverse values $k_y = 0, \pm \pi/2, \pi$, and for each we can allow an arbitrary ``Fermi segment,'' i.e., a contiguous region of occupied ${\bm k}$ orbitals.  For the $8 \times 4$ system, we optimized the trial energy over all distinct locations of these segments, the only restriction being the specified total filling, and found that only three bands are populated in a manner that respects the lattice symmetries.  For the $18 \times 4$ system, from the outset we restricted the optimization to such three-band states with inversion symmetry and found a state shown in Fig.~\ref{fig:VMCstate}.

\begin{figure}[t]
\centerline{\includegraphics[width=\columnwidth]{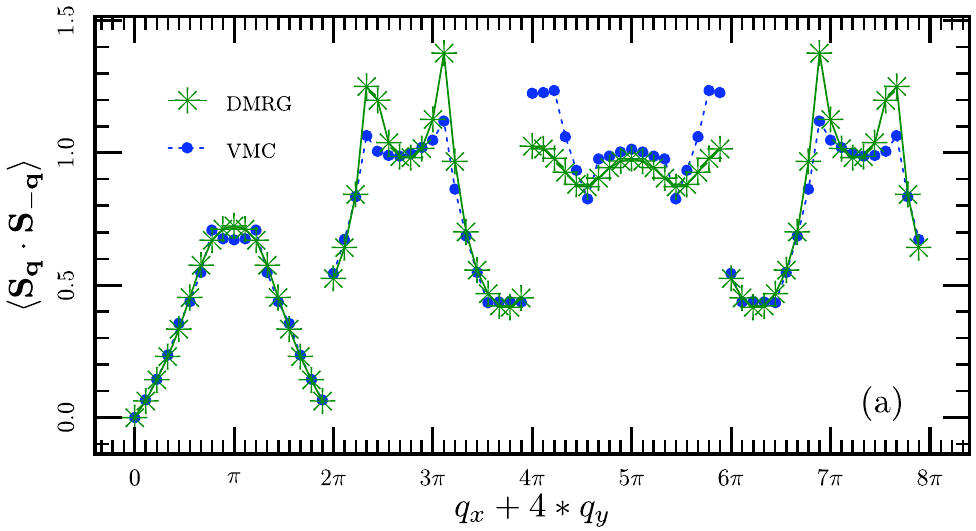}}
\centerline{\includegraphics[width=\columnwidth]{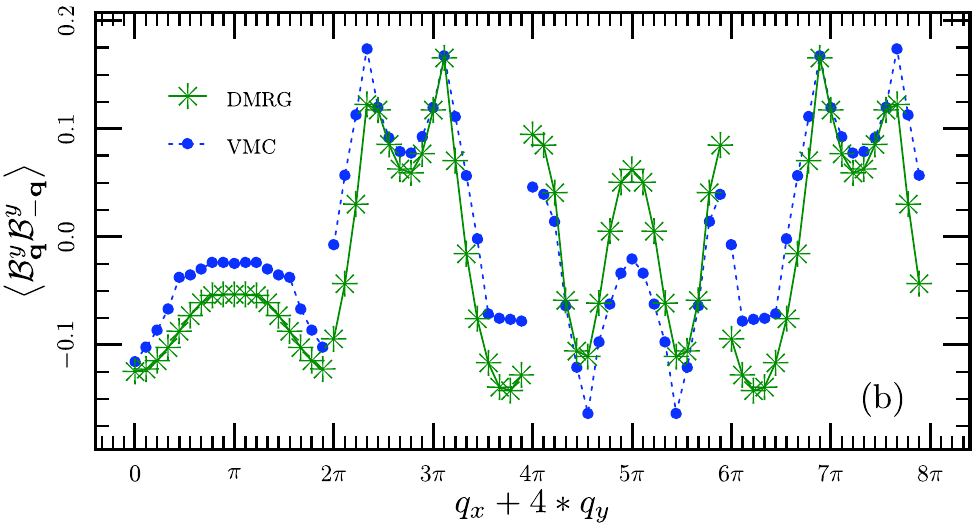}}
\centerline{\includegraphics[width=\columnwidth]{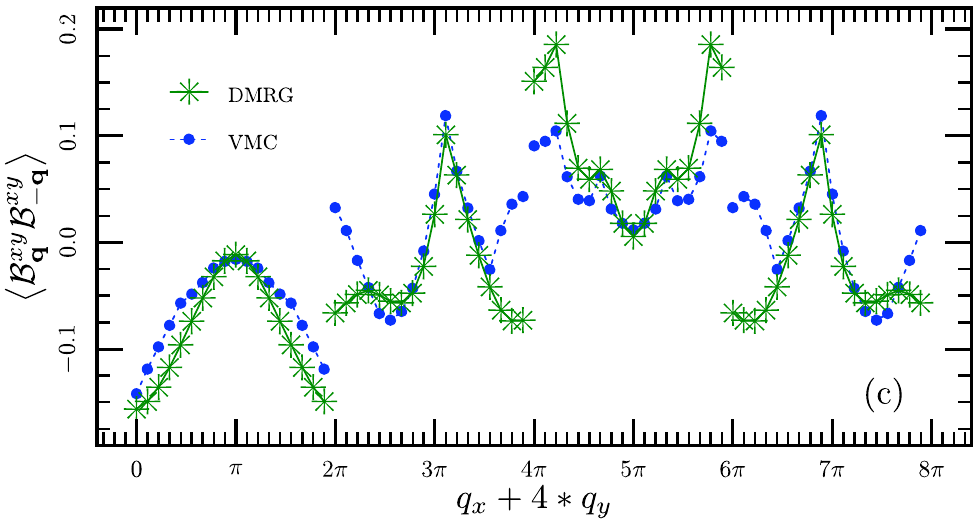}}
\caption{
(color online).  DMRG and VMC structure factors for the $18 \times 4$ triangular ladder at $K/J = 0.6$.  The horizontal axis is a linear representation of the $L \times 4$ mesh of $(q_x, q_y)$ points, and the disjoint curves from left to right correspond to $q_y = 0, \pi/2, \pi$, and $3\pi/2$. (a) Spin structure factor.  (b), (c) Dimer structure factor for bonds oriented in the $\hy$ direction and $\hx+\hy$ direction.
}
\label{fig:Sq}
\end{figure}

Figure~\ref{fig:Sq}(a) shows the spin structure factor measured in DMRG and calculated using the optimal VMC state on the $18 \times 4$ ladder, while Fig.~\ref{fig:Sq}(b) shows the dimer structure factor for bonds oriented in the $\hy$ direction.  In both figures, we see sharp peaks at wave vectors $(3 \times 2\pi/18, \pi/2)$ and $(10 \times 2\pi/18, \pi/2)$.  These are reproduced by the VMC wave function and the wave vectors can be associated with spinon transfers between right-mover and left-mover Fermi points differing by $\Delta k_y = \pm \pi/2$ in Fig.~\ref{fig:VMCstate}.

Figure~\ref{fig:Sq}(c) shows the dimer structure factor for bonds oriented in the $\hxy$ direction.  Here, we see notable peaks at wave vectors $\pm (2 \times 2\pi/18, \pi)$ and $\pm (6 \times 2\pi/18, \pi)$ in addition to some of those previously mentioned. The VMC agreement at these two points is again striking and can be associated with $\pm 2 k_F^{(-\pi/2)}$ and $\pm 2 k_F^{(+\pi/2)}$ spinon transfers in Fig.~\ref{fig:VMCstate}.  While there are quantitative discrepancies between the VMC and DMRG approaches, the overall agreement in the location of sharp features is notable. We do not show chirality structure factor results but can report that no ordering was observed over the regime of parameters shown in our phase diagram.

A summary of the gauge field theory for the SBM-3 phase is as follows.  In the mean field, there are three partially filled bands for each spinon species as in Fig.~\ref{fig:VMCstate}.  Linearizing near the Fermi points and bosonizing, there are total of six modes: three in the ``charge'' sector and three in the ``spin'' sector.  In the quasi-1D system, the gauge fluctuations beyond the mean field eliminate the overall charge mode, leaving five gapless modes.  In principle, one should be able to verify this prediction by analyzing the entanglement entropy with DMRG and extracting the central charge.  Indeed, we were able to confirm this for an $8 \times 4$ system, but exhausted the DMRG resources for larger sizes.

The three modes in the spin sector cannot have nontrivial Luttinger parameters due to the SU(2) invariance, while there are two Luttinger parameters describing the charge sector. Inspection of all allowed interactions shows that the SBM-3 can in principle be a stable phase, although there are many channels where this multimode system can become unstable.  Dominant spin and dimer correlations are expected at wave vectors $\pm 2{\bm k}_{Fa}$, where ${\bm k}_{Fa}$ runs over the three right-mover Fermi points in Fig.~\ref{fig:VMCstate}.  Potentially enhanced correlations are also expected at wave vectors $\pm ({\bm k}_{Fa} + {\bm k}_{Fb})$; at each such wave vector, the ``multiplet'' of observables with the same power law contains spin, dimer, scalar chirality, and vector chirality. The Gutzwiller wave functions correspond to a fine-tuned theory where all Luttinger parameters are equal to unity and the above wave vectors show the same power law $x^{-5/3}$.

While the DMRG does not show some of the expected wave vectors, the overall match with the VMC suggests that this may be due to matrix element effects.  It can also be that the SBM-3 is eventually unstable here, but it clearly is a good starting point for understanding the remarkable phase found in the Heisenberg plus ring model.

%%%%%%%%%%%%%%%%%%%%%%%%%%%%%%%%%%%%%%%%%%%%%%%%%%%%%%%%%%%%%%%%%%%%%%%
{\it Interpolation between the triangular and square limits.}---Motivated by relatives of the \ET\ and \dmit\ spin-liquid materials, we extend the study by allowing a different diagonal coupling $J_\hxy < J_\hx = J_\hy$.  We also vary $K_{\hx, \hy}$, while the remaining ring couplings are fixed by
\begin{equation}
K_{\hx, \hxy} = K_{\hy, \hxy} = \frac{J_\hxy}{J_\hx} K_{\hx, \hy}  ~.
\label{scaledKs}
\end{equation}
In an anisotropic electronic system with hoppings $t_{\hat{a}}$ in the $\hat{a}$ direction, the Heisenberg couplings are $J_{\hat{a}} \sim t_{\hat{a}}^2/U$ and the ring couplings $K_{\hat{a}, \hat{b}} \sim t_{\hat{a}}^2 t_{\hat{b}}^2 / U^3$, so their ``anisotropies'' are indeed related: $K_{\hat{a}, \hat{b}} \sim J_{\hat{a}} J_{\hat{b}} / U$.

Figure~\ref{fig:phased} gives the phase diagram in the $K_{\hx, \hy}/J_\hx$---$J_\hxy/J_\hx$ plane determined from the DMRG and VMC.  We see three prominent phases.  Along the $K_{\hx, \hy} = 0$ axis, the system is in the rung phase.  Going to the square limit, $J_\hxy = 0$, and then increasing $K_{\hx,\hy}$, the system undergoes a transition to a staggered dimer phase for $K_{\hx, \hy} \geq 0.8$.  This agrees with an earlier study \cite{Lauchli05} of the 2D square lattice model with ring exchanges that found staggered dimer phase in the same regime.  In the present four-leg system, the dimers orient transverse to the ladder ($\hy$ direction).  As $J_\hxy$ increases towards the triangular regime, the dimer phase expands to smaller values of $K_{\hx, \hy}$.

Significantly, the dimer phase disappears for anisotropy $0.8 \leq J_\hxy \leq 1$ and moderate to large ring exchange.  Here the DMRG finds a spin-liquid state that fits with the SBM-3.  The dimer phase touches the triangular axis near $K = 0.2$-$0.25$ where it becomes degenerate with the VBS state found earlier, cf.~Figs.~\ref{fig:VBS}(a)~and~\ref{fig:VBS}(b).

%%%%%%%%%%%%%%%%%%%%%%%%%%%%%%%%%%%%%%%%%%%%%%%%%%%%%%%%%%%%%%%%%%%%%%%
{\it Discussion.}---The four-leg ladder captures a good deal of local physics of the 2D model and allows guesses about the 2D phase diagram. 
First, in the rung phase near the square limit we observe strong spin correlations at $(\pi,\pi)$ and can view this region as a ladder descendant of the square lattice Neel state.
Second, the staggered $\hy$ dimer state is a descendant of the 2D staggered dimer state.  In 2D there is a degeneracy between cases where the strongest bonds are oriented in the $\hy$ or $\hx$ direction.  Third, the four-leg SBM-3 state is a direct descendant of the 2D SBM.
To summarize, we expect the 2D Neel, staggered VBS, and spin-liquid phases to occupy roughly similar regions as the rung, staggered $\hy$ dimer, and SBM-3 phases in Fig.~\ref{fig:phased}.
We do not venture to speculate how the three phases meet, particularly since additional phases enter into competition.  Specifically, the 2D triangular lattice with $K \leq 0.1$ has the 120$^\circ$ AF phase \cite{LiMing}.  Also, series expansions \cite{Weihong99} suggest that the Heisenberg model ($K=0$ axis) has a columnar VBS phase for $0.7 < J_\hxy/J_\hx < 0.9$, which is different from the staggered VBS stabilized by the ring exchanges.

Naive Hubbard model estimates for the organic spin-liquid materials suggest that they lie in the challenging regime of strong frustration ($J_\hxy \sim J_\hx \sim J_\hy$) and small to intermediate $K \sim 0.2 J$.  Our study shows that the SBM phase is a viable contender.  More realistic treatments such as inclusion of further ring exchanges (pursued systematically for the Hubbard model in \cite{Yang2010}) and long-range Coulomb interactions may tilt the balance towards the spin-liquid phase and deserve further study.

%%%%%%%%%%%%%%%%%%%%%%%%%%%%%%%%%%%%%%%%%%%%%%%%%%%%%%%%%%%%%%%%%%%%%%%
This work was supported by NSF Grants No. DMR-0529399 (M.S.B.\ and M.P.A.F.), DMR-0611562, 0906816 and MRI-0958596 (D.N.S.), DMR-0907145 (O.I.M), and the A. P. Sloan Foundation (O.I.M.).

\end{document}